\documentclass[iop,revtex4.1,numberedappendix,appendixfloats]{emulateapj}

\usepackage{natbib}
\usepackage{amsmath}
\usepackage{color}
\usepackage{url}
\usepackage{ulem}
\usepackage{multirow}
\usepackage{amsmath}
\usepackage{wasysym}

\usepackage{rotating, graphicx}

\bibpunct{(}{)}{;}{a}{}{,}

\def\apjl{ApJ}%
\def\apjs{ApJS}%
\def\ao{Appl.~Opt.}%
\def\aap{A\&A}%
\shorttitle{The destruction and recreation of the X-ray corona in a changing-look AGN}
\shortauthors{Ricci et al.}

\begin{document}

\title{The destruction and recreation of the X-ray corona in a changing-look \newline Active Galactic Nucleus}

\author{C. Ricci\altaffilmark{1,2,3,*}, E. Kara\altaffilmark{4}, M. Loewenstein\altaffilmark{5,6}, B. Trakhtenbrot\altaffilmark{7}, I. Arcavi\altaffilmark{7,8}, R. Remillard\altaffilmark{4}, A. C. Fabian\altaffilmark{9}, K. C. Gendreau\altaffilmark{5}, Z. Arzoumanian\altaffilmark{5}, R. Li\altaffilmark{2}, L. C. Ho\altaffilmark{2,10}, C. L. MacLeod\altaffilmark{11}, E. Cackett\altaffilmark{12}, D. Altamirano\altaffilmark{13}, P. Gandhi\altaffilmark{13}, P. Kosec\altaffilmark{9}, D. Pasham\altaffilmark{4}, J. Steiner\altaffilmark{4}, C.-H. Chan\altaffilmark{14,7}
}

\altaffiltext{1}{N\'ucleo de Astronom\'ia de la Facultad de Ingenier\'ia, Universidad Diego Portales, Av. Ej\'ercito Libertador 441, Santiago, Chile} 
\altaffiltext{2}{Kavli Institute for Astronomy and Astrophysics, Peking University, Beijing 100871, China}
\altaffiltext{3}{George Mason University, Department of Physics \& Astronomy, MS 3F3, 4400 University Drive, Fairfax, VA 22030, USA}
\altaffiltext{4}{MIT Kavli Institute for Astrophysics and Space Research, 70 Vassar Street, Cambridge, MA 02139, USA}
\altaffiltext{5}{Astrophysics Science Division, NASA Goddard Space Flight Center, 8800 Greenbelt Road, Greenbelt, MD 20771, USA}
\altaffiltext{6}{Department of Astronomy, University of Maryland, College Park, MD 20742, USA}
\altaffiltext{7}{School of Physics and Astronomy, Tel Aviv University, Tel Aviv 69978, Israel}
\altaffiltext{8}{CIFAR Azrieli Global Scholars program, CIFAR, Toronto, Canada}
\altaffiltext{9}{Institute of Astronomy, University of Cambridge, Madingley Road, CB3 0HA Cambridge, UK}
\altaffiltext{10}{Department of Astronomy, School of Physics, Peking University, Beijing 100871, China}
\altaffiltext{11}{Center for Astrophysics, Harvard \& Smithsonian, 60 Garden Street, Cambridge, MA 02138-1516, USA}
\altaffiltext{12}{Department of Physics \& Astronomy, Wayne State University, 666 West Hancock Street, Detroit, MI 48201, USA}
\altaffiltext{13}{Department of Physics \& Astronomy, University of Southampton, Southampton, Hampshire S017 1BJ, UK}
\altaffiltext{14}{Racah Institute of Physics, Hebrew University of Jerusalem, Jerusalem 91904, Israel}

\altaffiltext{*}{claudio.ricci@mail.udp.cl}

\begin{abstract}

We present the drastic transformation of the X-ray properties of the active galactic nucleus 1ES\,1927+654, following a changing-look event. After the optical/UV outburst the power-law component, produced in the X-ray corona, disappeared, and the spectrum of 1ES\,1927+65 instead became dominated by a blackbody component ($kT\sim 80-120$\,eV). This implies that the X-ray corona, ubiquitously found in AGN, was destroyed in the event. 
Our dense $\sim 450$\,day long X-ray monitoring shows that the source is extremely variable in the X-ray band. On long time scales the source varies up to $\sim 4$\,\,dex in $\sim 100$\,days, while on short timescales up to $\sim2$\,\,dex in $\sim 8$ hours. The luminosity of the source is found to first show a strong dip down to $\sim 10^{40}\rm\,erg\,s^{-1}$, and then a constant increase in luminosity to levels exceeding the pre-outburst level $\gtrsim $300 days after the optical event detection, rising up asymptotically to $\sim 2\times10^{44}\rm\,erg\,s^{-1}$. As the X-ray luminosity of the source increases, the X-ray corona is recreated, and a very steep power-law component ($\Gamma\simeq 3$) reappears, and dominates the emission for 0.3--2\,keV luminosities $\gtrsim 10^{43.7}\rm\,erg\,s^{-1}$, $\sim 300$\,days after the beginning of the event. We discuss possible origins of this event, and speculate that our observations could be explained by the interaction between the accretion flow and debris from a tidally disrupted star. Our results show that changing-look events can be associated with dramatic and rapid transformations of the innermost regions of accreting SMBHs.

\end{abstract}


\section{Introduction}

\setcounter{footnote}{0}

\noindent Accreting supermassive black holes (SMBHs) are known to show variable optical, ultraviolet and X-ray emission. One of the most intriguing aspects of this behavior is associated with ``changing-look' sources, in which the optical/ultraviolet broad emission lines, produced by rapidly-moving material surrounding the SMBH (e.g., \citealp{Kaspi:2000zy}), appear or disappear (e.g., \citealp{Shappee:2014mw,LaMassa:2015ne,Trakhtenbrot:2019qy}). This implies that changing-look active galactic nuclei (AGN) transition from type\,1 (showing both broad and narrow ionic emission lines in the optical) to type\,2 (showing only narrow lines), or vice versa. The physical mechanism producing these changing-look events is still hotly debated; the observed changes could be related to radiation pressure instabilities in the disk \citep{Sniegowska:2019jt}, to state transitions \citep{Noda:2018le,Ruan:2019wc}, or to variations in accretion rate \citep{Elitzur:2014rp}, possibly related to transient events \citep{Merloni:2015on}, with the source going through a highly-accreting phase and producing broad lines. 

\begin{figure*}[t!]
\centering
\includegraphics[width=0.8\textwidth]{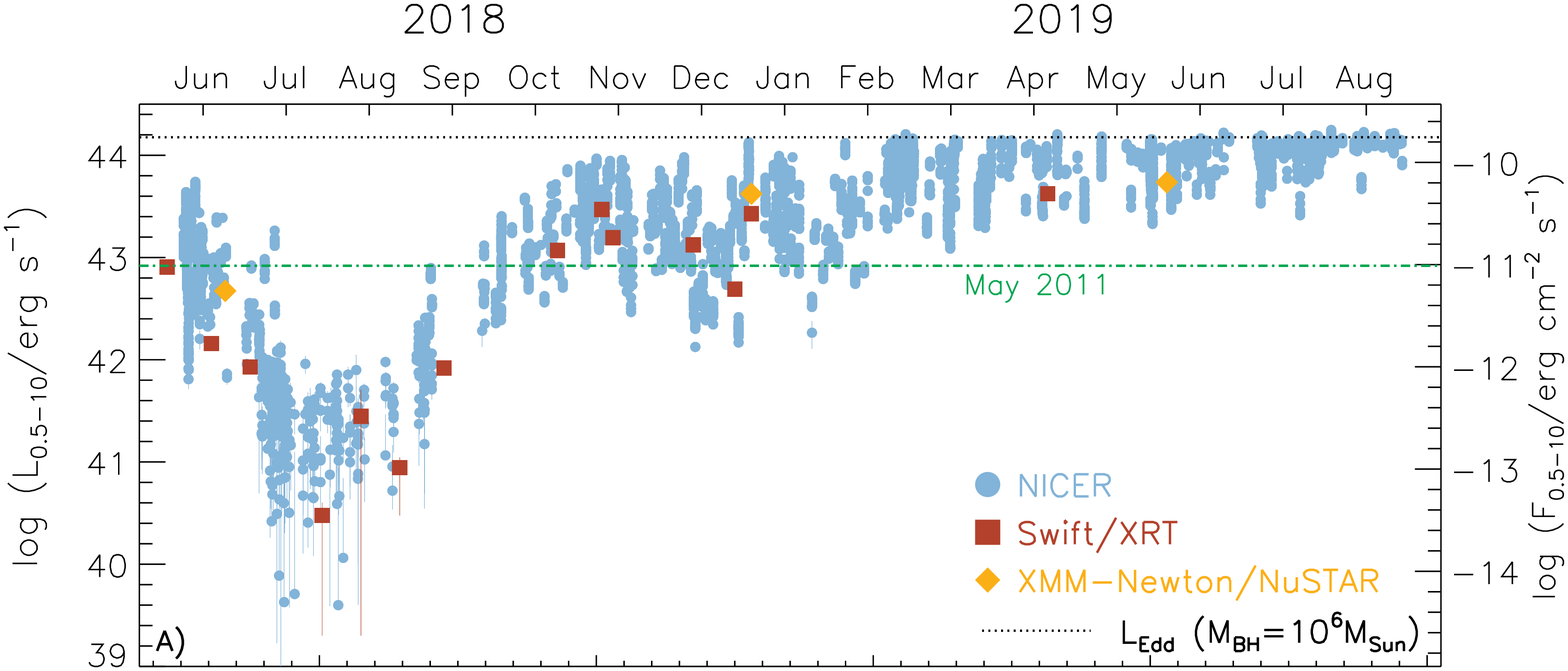}
\includegraphics[width=0.8\textwidth]{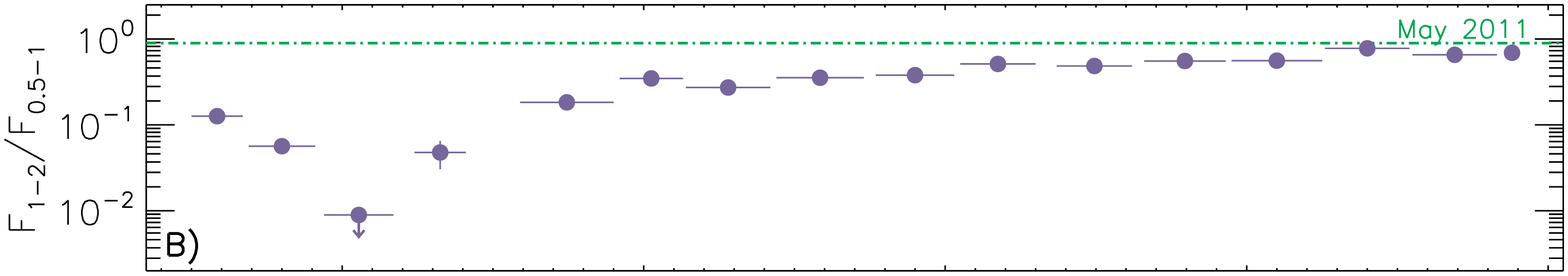}
\includegraphics[width=0.8\textwidth]{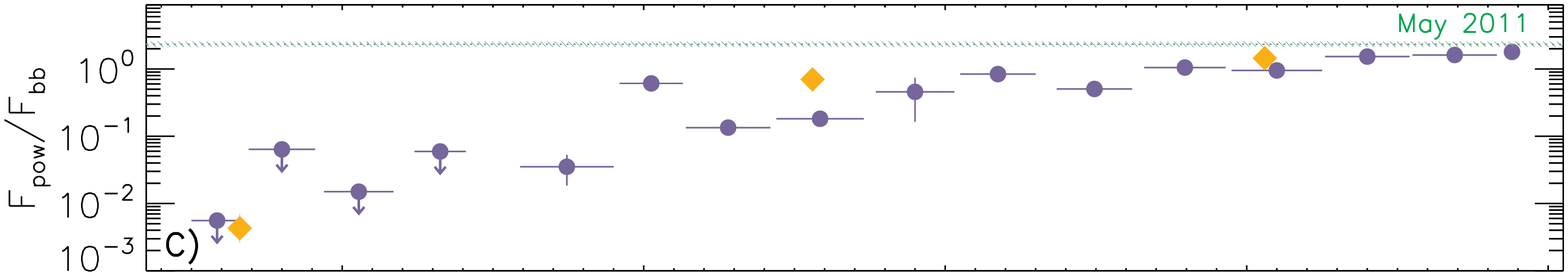}
\includegraphics[width=0.8\textwidth]{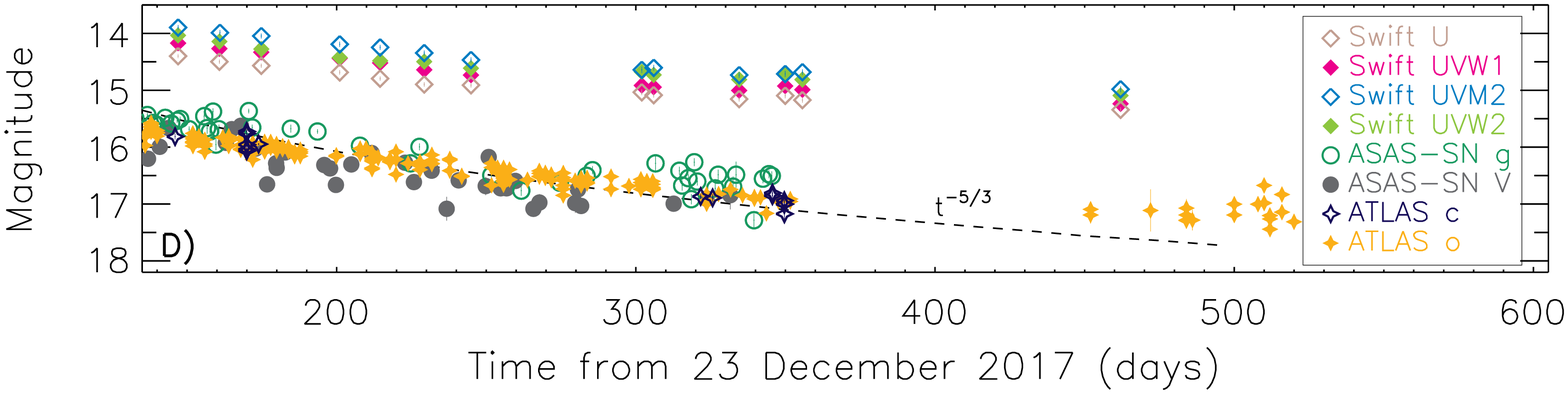}
% %
% %
 \caption{Long-term light curve of 1ES\,1927+654. {\it Panel A:} 0.5--10\,keV light curve, including {\it NICER} (cyan circles), {\it Swift}/XRT (red squares) and {\it XMM-Newton}/{\it NuSTAR} (yellow diamonds) data. Our first X-ray observation (May 17 2018) was carried out $\sim 150$ days after the detection of the outburst in the optical (December 23 2017; \citealp{Trakhtenbrot:2019qy}). The 0.5--10\,keV luminosity before the event, inferred by the 2011 {\it XMM-Newton} observation \citep{Gallo:2013hq}, was $8.3\times 10^{42}\rm\,erg\,s^{-1}$ (horizontal green dot-dashed line). The black dotted line represents the Eddington luminosity for $M_{\rm\,BH}= 10^{6}\,M_{\odot}$ (see Appendix\,\ref{sec:BHmass}). {\it Panel B:} Ratio between the 1--2\,keV and 0.5--1\,keV flux for the {\it NICER} observations, showing the softening after the event, followed by an hardening as the overall luminosity increased. {\it Panel C}: ratio between the power-law and blackbody flux in the 0.5--2\,keV band. {\it Panel D:} optical/UV lightcurve \citep{Trakhtenbrot:2019qy}, showing a very different behavior with respect to the X-ray band. The time of the event ($t_{0}$) adopted for the $t^{-5/3}$ curve (black dashed line) is December 23 2017.
 }
 \label{fig:longtermLC}
\end{figure*}

So far only a few dozens of these systems have been discovered (e.g., \citealp{Yang:2018qt,MacLeod:2019pm}) and, due to their rarity, it has been very difficult to catch any of them during the transition phase. We were recently able to observe, for the first time, an AGN in the act of changing phase. 
1ES\,1927+654 ($z=0.019422$, \citealp{Trakhtenbrot:2019qy}) was previously classified as a type\,2 AGN (e.g., \citealp{Gallo:2013hq} and references therein). In March 2018 the All-Sky Automated Survey for SuperNovae \citep{Shappee:2014mw} showed that the source suddenly increased in flux by an order of magnitude in the optical V-band (ASASSN-18el/AT2018zf,  \citealp{Nicholls:2018yo}). The analysis of Asteroid Terrestrial-impact Last Alert System (ATLAS, \citealp{Tonry:2018qy}) data indicated that the event was first detected on December 23 2017. An optical spectroscopic follow-up campaign showed first a very blue continuum and then, $1-3$ months after the optical flux rise, evidence of strong broad Balmer emission lines \citep{Trakhtenbrot:2019qy}. 

Right after the appearance of the broad optical lines we started monitoring the source in the X-ray band. X-rays in AGN are primarily created by Comptonization of optical/UV disk photons in a corona of hot electrons, located within a light-hour from the accreting system \citep{Fabian:2009bz}, and likely powered by the magnetic field of the accretion disk \citep{Merloni:2001qy}. These electrons transfer their energy to the disk photons, up-scattering them into the X-rays \citep{Haardt:1991dq}, and creating the power-law continuum ubiquitously found in AGN. X-ray observations therefore provide an extremely powerful tool to study the inner regions of AGN and the physical conditions of the accretion flow. In this work we discuss the peculiar X-ray properties of 1ES\,1927+654 after the changing-look event.
Throughout the paper we adopt standard cosmological parameters ($H_{0}=70\rm\,km\,s^{-1}\,Mpc^{-1}$, $\Omega_{\mathrm{m}}=0.3$, $\Omega_{\Lambda}=0.7$). Unless otherwise stated, uncertainties are quoted at the 90\% confidence level.

\section{The disappearance and reappearance of the power-law component}

Our  X-ray campaign consists of 265 {\it NICER} \citep{Gendreau:2012cr,Arzoumanian:2014qf} observations (for a total of 678ks), 14 {\it Neil Gehrels Swift Observatory} \citep{Gehrels:2004dq} observations (26 ks), and three joint {\it XMM-Newton}/{\it NuSTAR} \citep{Jansen:2001ve,Harrison:2013zr} observations (158/169 ks). Details about the data reduction and spectral analysis of all these X-ray observations are reported in a dedicated companion publication (\citealp{Ricci:2020}, see also \citealp{Kara:2018fc}), while here we focus on the unexpected spectral properties of 1ES\,1927+654 after the changing-look event.

\smallskip

Our monitoring campaign shows that the source is extremely variable in the 0.5--10\,\,keV band (panel\,\,A of Fig.\,\ref{fig:longtermLC}), with very strong intraday variability (up to $\sim$2\,\,dex in $\sim$8 hours). The strongest variability is however found on longer time scales. The first X-ray observations, carried out in May 2018, found 1ES\,1927+654 at a similar luminosity level of the previous {\it XMM-Newton} observation ($L_{0.5-10}=8.3\times 10^{42}\rm\,erg\,s^{-1}$, \citealp{Gallo:2013hq}; green dot-dashed line in panel\,\,A of Fig.\,\ref{fig:longtermLC}). At $t\sim 160$\,\,days from the optical event detection the luminosity of the source started to decrease rapidly, reaching $L_{0.5-10}\sim 10^{40}\rm\,erg\,s^{-1}$ at $t\sim 200$\,days. After this dip we observed a constant increase in luminosity, to levels exceeding the pre-outburst value at $t\gtrsim$300 days. The luminosity then rose up asymptotically to $\sim 2\times10^{44}\rm\,erg\,s^{-1}$, with the intraday variability decreasing to $\sim 0.4$\,dex. The X-ray flux variability is tightly connected with the spectral variability, with the source showing a clear harder-when-brighter behavior (panels\,\,B and C of Fig.\,\ref{fig:longtermLC}). Interestingly, the X-ray and optical/UV fluxes are completely disconnected, with the latter showing a monotonic declining trend for $t> 150$\,\,days (Panel\,\,D and \citealp{Trakhtenbrot:2019qy}).

\begin{figure}[t!]
 \centering
 \includegraphics[width=0.48\textwidth]{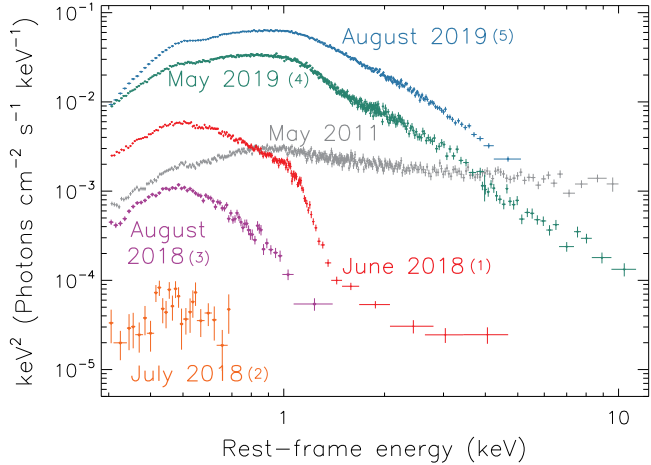}
 \caption{X-ray spectral evolution of 1ES\,1927+654. Spectra of the May\,2011 (gray points;  {\it XMM-Newton} EPIC/PN), June\,\,2018 (red points;  {\it XMM-Newton} EPIC/PN and MOS), July\,\,2018 (orange points;  {\it NICER}), August\,\,2018 (purple points;  {\it NICER}), May\,\,2019 (green points;  {\it XMM-Newton} EPIC/PN and {\it NuSTAR}) and July\,\,2019 (blue points;  {\it NICER}) observations. Soon after the event (June--August\,\,2018) the X-ray spectrum of 1ES\,1927+654 was extremely soft, and did not show the strong hard X-ray component ubiquitously found in AGN and present during the 2011 observation. In the observations carried out in 2019 the flux of the source is higher and a steep power-law component has reappeared.
 }
\label{fig:spectra}
\end{figure}

Previous X-ray observations \citep{Gallo:2013hq} of 1ES\,1927+654 (gray spectrum in Fig.\,\ref{fig:spectra}) found that the source is unobscured ($N_{\rm H}\simeq 10^{20}\rm\,cm^{-2}$), and that its X-ray spectrum could be well represented by a model that includes a power law ($\Gamma=2.39\pm0.04$) and a blackbody component ($kT=170\pm5$\,eV). 
After the optical brightening event, 1ES\,1927+654 showed a spectrum completely different from the 2011 {\it XMM-Newton} observation. In the first {\it XMM-Newton} observation of our campaign (June\,\,2018, red spectrum in Fig.\,\ref{fig:spectra}), the spectrum was extremely soft, with the emission being dominated by a blackbody-like component. The disappearance of the power-law component, which has not hitherto been observed in AGN, implies that the dramatic event that created the broad lines also destroyed the X-ray corona.
The X-ray spectra of 1ES\,1927+654 can be typically well modeled by a blackbody, two Gaussian emission lines and a power-law component. Absorption from outflowing ionized gas (with $N_{\rm H}\sim 10^{20}\rm\,cm^{-2}$) is also clearly observed in the {\it XMM-Newton}/RGS observations (see \citealp{Ricci:2020}).
During the June\,\,2018 {\it XMM-Newton} observation the blackbody was found to have a temperature of $kT=102\pm1$\,eV, and only a very faint power-law emission was detected at $E\gtrsim 1.5$\,keV, carrying a very small fraction ($0.4\pm0.3\%$) of the total 0.3--2\,keV flux. 
The power-law component is undetected or extremely faint also in the observations carried out by {\it NICER} in July\,\,2018, when the luminosity was at its lowest level, and in August\,\,2018 (spectra 2 and 3 in Fig.\,\ref{fig:spectra}, respectively). The spectral analysis of the stacked {\it NICER} spectra, obtained combining observations carried out in June to mid-August\,\,2018, recovered only upper limits for the ratio between the power-law and blackbody flux in the 0.3--2\,keV band. These upper limits range between $\lesssim 0.5$\% (June\,\,2018) and $\lesssim 6$\% (August\,\,2018). As the luminosity of the source increased above the pre-outburst level, our {\it XMM-Newton}/{\it NuSTAR} and {\it NICER} observations show that the power-law component reappeared (spectra 4 and 5). Interestingly, the photon index of this power-law component, $\Gamma\simeq 3$, is considerably higher than what is commonly found in AGN ($\Gamma\sim1.8$; e.g., \citealp{Ricci:2017fj}).

\begin{figure}[t!]
\begin{center}
\includegraphics[width=0.48\textwidth]{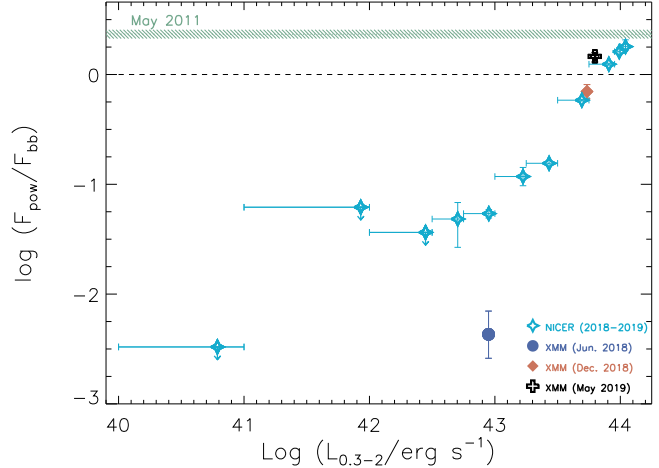}

    \caption{Ratio between the flux of the power-law ($F_{\rm pow}$) and blackbody ($F_{\rm bb}$) component versus the 0.3--2\,keV luminosity for the {\it NICER} (May\,\,2018 to August\,\,2019) and {\it XMM-Newton} observations. The horizontal black dashed line represents the $F_{\rm pow}=F_{\rm bb}$ case. The plot shows a clear positive relation between $F_{\rm pow}/F_{\rm bb}$ and the luminosity, with the spectrum becoming dominated by the power-law emission at $\log (L_{0.3-2}/\rm erg\,s^{-1})\simeq 43.7$, and reaching the level of $F_{\rm pow}/F_{\rm bb}$ observed in 2011 (green squares) at $\simeq 10^{44}\rm\,erg\,s^{-1}$.}
    \label{fig:bb_po_ratio}
  \end{center}
\end{figure}

In the X-ray observations carried out in late 2018 and in 2019 the fraction of the total 0.3--2\,keV flux ascribed to the power-law component is significantly higher than in June--August\,\,2018, and is found to dominate the X-ray emission in the May\,\,2019 ($F_{\rm pow}/F_{\rm bb}=1.45^{+0.13}_{-0.16}$) {\it XMM-Newton}/{\it NuSTAR} observation.  An increase with the luminosity of the ratio between the flux of the power law and that of the blackbody is clearly detected when stacking the {\it NICER} observations based on their 0.5--10\,keV luminosities (cyan stars in Fig.\,\ref{fig:bb_po_ratio}). At the lowest luminosities ($L_{0.5-10}\sim10^{40}-10^{41}\rm\,erg\,s^{-1}$) the ratio is $F_{\rm pow}/F_{\rm bb}\leq 0.003$, which is $\gtrsim 850$ times lower than the value found in 2011  ($F_{\rm pow}/F_{\rm bb}=2.6\pm0.2$), before the optical outburst. The power-law starts to dominate over the blackbody component for 0.3--2\,keV luminosities $\gtrsim 10^{43.7}\rm\,erg\,s^{-1}$, reaching a ratio between the power law and the blackbody flux similar to that found in the 2011 observations at $\sim 10^{44}\rm\,erg\,s^{-1}$. All this shows that, as the source becomes brighter, the X-ray corona is being recreated. Such a transition has not been observed before. This can also be observed by comparing the lightcurve to a model-independent flux ratio and to $F_{\rm pow}/F_{\rm bb}$ (panels\,\,B and C of Fig.\,\ref{fig:longtermLC}). Our spectral analysis of the two latest {\it XMM-Newton}/{\it NuSTAR} observations, which display a strong power-law component, shows that a cutoff at $E_{\rm C}\simeq 3$\,keV is needed to reproduce the data (see \citealp{Ricci:2020} for details). This cutoff energy is considerably lower than that observed in AGN ($E_{\rm C}\sim 50-300$\,keV; e.g., \citealp{Ricci:2017fj}), and points to a temperature of the corona of just $kT_{\rm\,e}\simeq 1.5$\,keV, assuming that the Comptonizing plasma is optically-thin. This assumption is supported by the fact that, considering $\Gamma$ and $E_{\rm C}$, and using Eq.\,6 of \citet{Ricci:2018mp}, the optical depth of the plasma would be $\tau\sim 0.15$. The low temperature and optical depth also suggest that the X-ray corona is in the process of being reformed, and could imply that the magnetic field powering it is still not strong enough to support the typical X-ray emitting plasma found in AGN.

\section{Discussion}

The surprisingly soft X-ray continuum (Fig.\,\ref{fig:spectra}), together with the dramatic X-ray variability (Fig.\,\ref{fig:longtermLC}), suggests that the source underwent some catastrophic event that restructured its accretion flow. The lack of X-ray flux above 2\,keV implies that the event strongly affected the X-ray corona (and possibly the innermost regions of the accretion flow), possibly completely destroying it. As the X-ray luminosity of 1ES\,1927+654 increases, the power-law component reappears, and starts to dominate the overall 0.3--2\,keV flux (Fig.\,\ref{fig:bb_po_ratio}). This suggests that the X-ray corona is in the process of being recreated.

It has been argued that changing-look events in AGN might be powered by hard-to-soft state transitions \citep{Noda:2018le,Ruan:2019wc}, analogous to those observed in black hole X-ray binaries, where the inner disk evaporates into an advection-dominated accretion flow. In this scenario the timescales for the evaporation of the accretion flow should be the viscous timescale. The very soft spectrum of 1ES\,1927+654 is reminiscent of the spectral shape of black hole X-ray binaries in the high/soft state (e.g., \citealp{Remillard:2006qh}), and the asymptotic behavior of the luminosity is similar to Eddington-limited accretion outbursts observed in stellar-mass systems (e.g., \citealp{Kimura:2016ue}). However, the timescales expected for a state transition (e.g., \citealp{Dunn:2010ak}) of a $10^6\,M_{\odot}$ black hole are considerably longer ($\sim 100$\,years) than what we observe in 1ES\,1927+654, where the spectral variations happen on timescales of hours. Strong magnetization \citep{Noda:2018le}, possibly coupled with geometrically-thick disks \citep{Dexter:2019rt}, could reduce the timescales, but would not be able to explain the spectral variability on $\lesssim1$\,day intervals detected by our {\it NICER} and {\it XMM-Newton} observations.

The event that destroyed the X-ray corona, while increasing the optical/UV flux and creating the broad Balmer lines, could be related to the tidal disruption of a star by the SMBH. 
While TDEs are rare (e.g., \citealp{Kawamuro:2016bs}), it has been argued that the presence of a long-lived accretion disk can enhance their rate with respect to inactive galaxies (e.g., \citealp{Karas:2007bs}). TDEs have been discussed as a possible driver of changing-look events \citep{Merloni:2015on}, and a few cases of AGN in which a TDE might have occurred have been reported over the past few years (e.g., \citealp{Liu:2020iz}). \citet{Merloni:2015on} found that the changing-look AGN SDSS\,0159+003 \citep{LaMassa:2015ne} displays a $t^{-5/3}$ behavior, which is typical of TDEs \citep{Rees:1988yk}, and argued that the event could be related to the disruption of a star by a $\sim 10^{8}\rm\,M_{\odot}$ SMBH. \citet{Campana:2015ki} discussed how the strong variability of IC\,3599 could be explained by multiple TDE events recurring every 9.5\,years. \citet{Blanchard:2017kb} proposed that the transient PS16dtm was due a TDE in a pre-existing narrow-line Seyfert 1 AGN with a $\sim 10^{6}\rm\,M_{\rm \odot}$ black hole. In this source the increase in the UV and optical flux was not accompanied by an increase of the X-ray flux, which was found to be at least an order of magnitude below the previous {\it XMM-Newton} detection. \citet{Blanchard:2017kb} argued that this might be due to absorption of the X-ray corona caused by the disk of debris.

\begin{figure}[t!]
\begin{center}
\includegraphics[height=0.48\textwidth,angle=270]{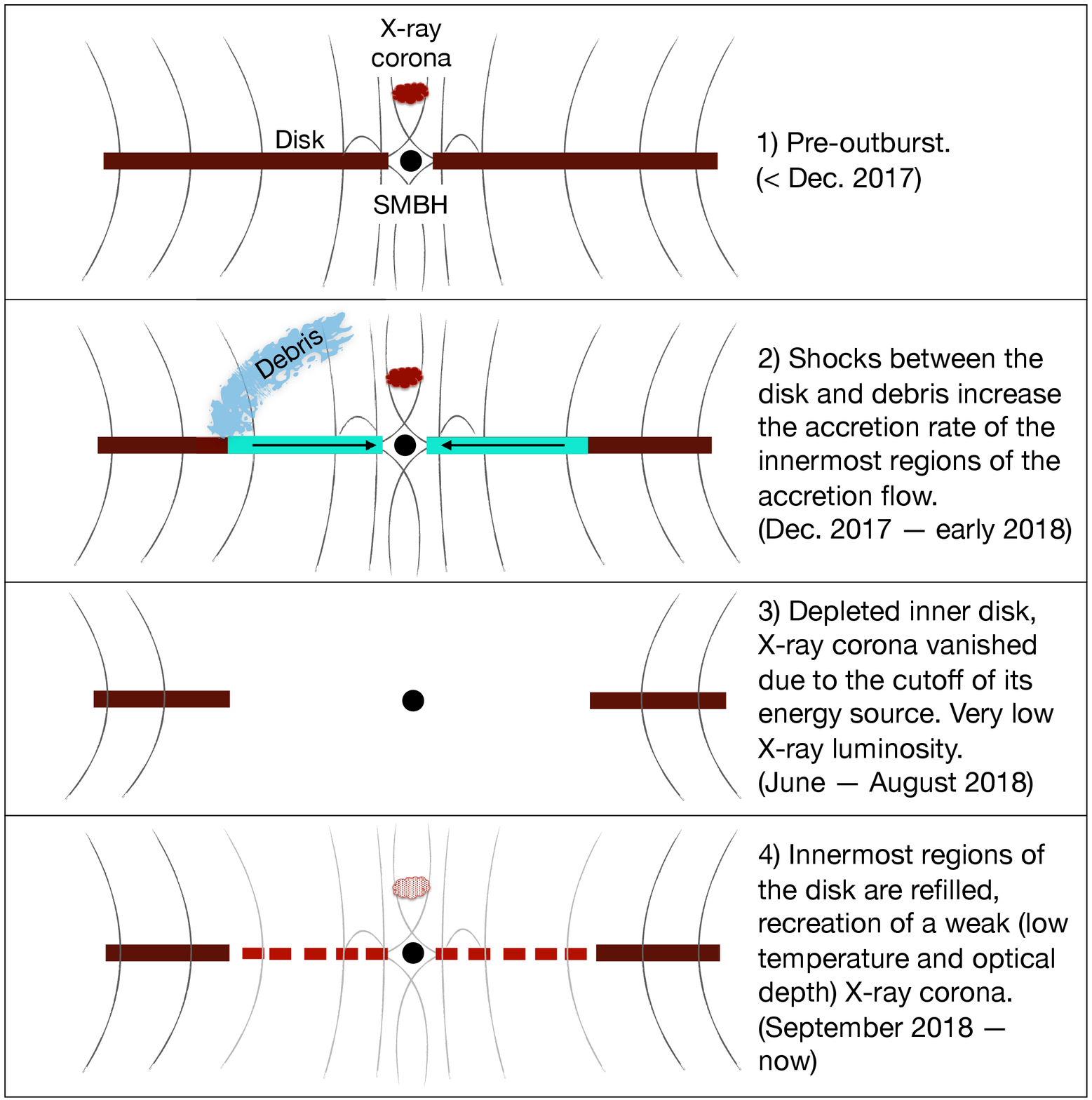}
    \caption{Schematic representation of the TDE scenario. The innermost regions of the accretion disk, where the magnetic field (black lines) powers the X-ray corona (1), are strongly affected by the interaction with the debris (2). The shocks excited by the debris increase the accretion rate in the inner disk, leading to a depletion of the accretion flow (3). Due to the disappearance of the inner disk and of its magnetic field, the X-ray corona is destroyed. As the disk is replenished the magnetic field starts reforming, together with the X-ray corona (4). }
    \label{fig:cartoon}
  \end{center}
\end{figure}

The very soft X-ray emission observed after the event, and the fact that the optical/UV flux decreases following a trend consistent with the $t^{-5/3}$ behavior \citep{Trakhtenbrot:2019qy} are in agreement with the TDE scenario (e.g., \citealp{Komossa:2015iz,Auchettl:2017xo}), while the re-brightening observed in the X-rays is very different from what is typically observed in TDEs. Interestingly, the behavior seen in 1ES\,1927+654 might be consistent with recent hydrodynamic simulations of a debris stream colliding with an accretion disk \citep{Chan:2019dx}. In this scenario, schematically illustrated in Fig.\,\,\ref{fig:cartoon}, the innermost regions of the accretion flow, where the X-ray coronal power-law component is produced, accrete faster than the outer parts, due to the efficient removal of angular momentum by shocks excited in the disk by the debris. The enhanced accretion rate empties the inner disk, which would lead to the destruction of the magnetic field pattern and cutting off the energy source of the X-ray corona. The disruption of the corona leads to the disappearance of the power law component and to a drastic decrease of the X-ray luminosity of the system. The non-steady accretion flow and/or the shocks could be responsible for the soft component observed below 2\,keV (see also \citealp{Ricci:2020}). The lack of UV broad lines \citep{Trakhtenbrot:2019qy} with ionization energy $\chi_{\rm ion}\gtrsim 15$\,eV, such as Mg\,\textsc{ii\,$\lambda$2798} ($\chi_{\rm ion}=15.04$\,eV) and C\,\textsc{iv\,$\lambda$1549} ($\chi_{\rm ion}=64.49$\,eV), also suggests a depletion of the inner disk. 
As the inner disk is replenished by material flowing in from the outer disk, the magnetic field around the SMBH is reformed, and restarts powering the X-ray corona; this leads to the reappearance of the power-law component in the X-ray spectrum. During this period an increase in the overall X-ray luminosity is expected, consistent with our observations. The asymptotic luminosity limit at $\sim1.5\times$\,10$^{44}$\,erg\,s$^{-1}$ might indicate that, in this framework, the replenishment of the disk is Eddington-limited (Fig.\,\ref{fig:spectra}). Interestingly, the observed decrease of the X-ray flux in PS16dtm after the event is similar to what we found for 1ES\,1927+654 during the first months after the event, although no re-brightening has been observed so far in this source. 1ES\,1927+654 also shares some observational properties with super-soft AGN (e.g., \citealp{Terashima:2012nx,Miniutti:2019rg,Giustini:2020qg}), objects that could be associated to TDEs (e.g., \citealp{Shu:2018bh,King:2020dm}), such as the harder-when-brighter behavior (see \citealp{Ricci:2020} for details), and a very prominent blackbody component, in particular during the lower luminosity intervals.

\section{Summary and conclusions}

1ES\,1927+654 is a type-2 AGN that was recently found to develop broad optical emission lines after an optical outburst (23 December 2017; \citealp{Trakhtenbrot:2019qy}). Following the changing-look event, we started an intense 450-day X-ray monitoring campaign, which includes observations from {\it NICER} (265 pointings), {\it Swift} (14) and {\it XMM-Newton}/{\it NuSTAR} (3). Our observations show that: 
\begin{itemize}
\item The source is extremely variable both on long and short time scales (see Fig.\,\ref{fig:longtermLC}), with the flux changing up to $\sim 4$\,\,dex in $\sim 100$\,days, and up to $\sim2$\,\,dex in $\sim 8$ hours. The luminosity of the source is found to first show a strong dip down to $\sim 10^{40}\rm\,erg\,s^{-1}$, and then a constant increase in luminosity to levels exceeding the pre-outburst level $\gtrsim$300 days after the optical event detection, rising up asymptotically to $\sim 2\times10^{44}\rm\,erg\,s^{-1}$. 
\item After the optical outburst the power-law component produced in the X-ray corona, which is ubiquitously observed in AGN, was found to have almost completely disappeared (Fig.\,\ref{fig:spectra}). The X-ray spectrum was instead dominated by a very strong blackbody component with $kT\simeq 100$\,eV. During the first (May--August\,\,2018) observations only a very faint power-law emission was detected at $E\gtrsim 1.5$\,keV, carrying only $\lesssim 1\%$ of the total 0.3--2\,keV flux. 
\item A clear positive correlation between the ratio of the power-law and blackbody flux ($F_{\rm pow}/F_{\rm bb}$) in the 0.3--2\,keV range and the luminosity is found stacking the {\it NICER} spectra (Fig.\,\ref{fig:bb_po_ratio}). At the lowest luminosities ($L_{0.5-10}\sim10^{40}-10^{41}\rm\,erg\,s^{-1}$) the ratio is $F_{\rm pow}/F_{\rm bb}\leq 0.003$, i.e. $\gtrsim 850$ times lower than the value found in 2011  ($F_{\rm pow}/F_{\rm bb}=2.6\pm0.2$), before the optical outburst. As the X-ray luminosity of the source increases, the X-ray corona is recreated, and a very soft power-law component ($\Gamma\simeq 3$) reappears, and dominates the emission for 0.3--2\,keV luminosities $\gtrsim 10^{43.7}\rm\,erg\,s^{-1}$, $\sim 300$\,days after the beginning of the event. 
\end{itemize}

These results show that changing-look events can be associated with dramatic and rapid transformations of the innermost regions of accreting SMBHs. We speculate that the behavior of 1ES\,1927+654 could be caused by the tidal disruption of a star by the accreting black hole, which is expected to deplete the innermost regions of the accretion flow \citep{Chan:2019dx}, therefore disrupting the magnetic field powering the X-ray corona (Fig.\,\ref{fig:cartoon}). As the inner regions of the accretion flows are replenished, the magnetic field around the SMBH is reformed, and restarts powering the X-ray corona. This would lead to the reappearance of the power-law component and to the increase of the X-ray luminosity.

Our coordinated multi-wavelength observations of the dramatic transformation of 1ES\,1927+654 offer an unprecedented opportunity to probe accretion physics, and to understand the relation between changing-look events and the innermost regions of the accreting system. Upcoming surveys, such as those that are or will be carried out by {\it eROSITA} \citep{Merloni:2012be}, the {\it Einstein Probe} \citep{Yuan:2015vg} and SDSS-V \citep{Kollmeier:2017uv}, are designed to reveal large samples of similar events, allowing to further constrain the typical timescales and the occurrence rates of such transformative events, and to understand their importance to massive black hole growth

\acknowledgments

We thank the referee for their very quick review, which helped us improve the quality of the manuscript. We acknowledge {\it XMM-Newton}, {\it NuSTAR} and {\it Swift} for the DDT observations they kindly guaranteed us. LH acknowledges financial support from the National Key R\&D Program of China grant No. 2016YFA0400702, and the National Science Foundation of China grants No. 11473002 and 1721303. CR acknowledges support from the Fondecyt Iniciacion grant 11190831. BT acknowledges support from the Israel Science Foundation (grant No. 1849/19). IA is a CIFAR Azrieli Global Scholar in the Gravity and the Extreme Universe Program and acknowledges support from that program, from the Israel Science Foundation (grant numbers 2108/18 and 2752/19), from the United States - Israel Binational Science Foundation (BSF), and from the Israeli Council for Higher Education Alon Fellowship. DA acknowledges support from the Royal Society. CHC is supported by ERC advanced grant ``TReX" and the CHE-ISF Center for Excellence in Astrophysics. PG acknowledges support from STFC and a UGC-UKIERI Thematic Partnership. CR acknowledges F. Bauer, C.S. Chang and the Santiago AGN community for useful discussion. Based on observations with the NASA/ESA/CSA Hubble Space Telescope obtained [from the Data Archive] at the Space Telescope Science Institute, which is operated by the Association of Universities for Research in Astronomy, Incorporated, under NASA contract NAS5-26555. Support for Cycle 25 Program GO-15604 was provided through a grant from the STScI under NASA contract NAS5-26555.

{\it Facilities:} \facility{NICER}, \facility{Swift}, \facility{XMM-Newton}, \facility{NuSTAR}.

\bibliographystyle{apj} %% 
\bibliography{1ES1927_xray.bib}

\appendix

\section{Black hole mass and Eddington ratio}\label{sec:BHmass}

A black hole mass of $1.9\times10^{7}\rm\,M_{\odot}$ has been inferred by considering the width of the broad Balmer lines \citep{Trakhtenbrot:2019qy}. However, due to the transient nature of the event, the clouds emitting the broad Balmer lines might not have had the time to virialize, as shown by their variable widths \citep{Trakhtenbrot:2019qy}. The stellar mass of the galaxy inferred from K-band photometry (e.g., \citealp{Kormendy:2013uf}) suggests a lower black hole mass ($M_{\rm BH}\sim 10^{6}M_{\odot}$; Li et al. in preparation). 

\begin{figure}[h!]
 \centering
\includegraphics[height=0.65\textwidth,angle=270]{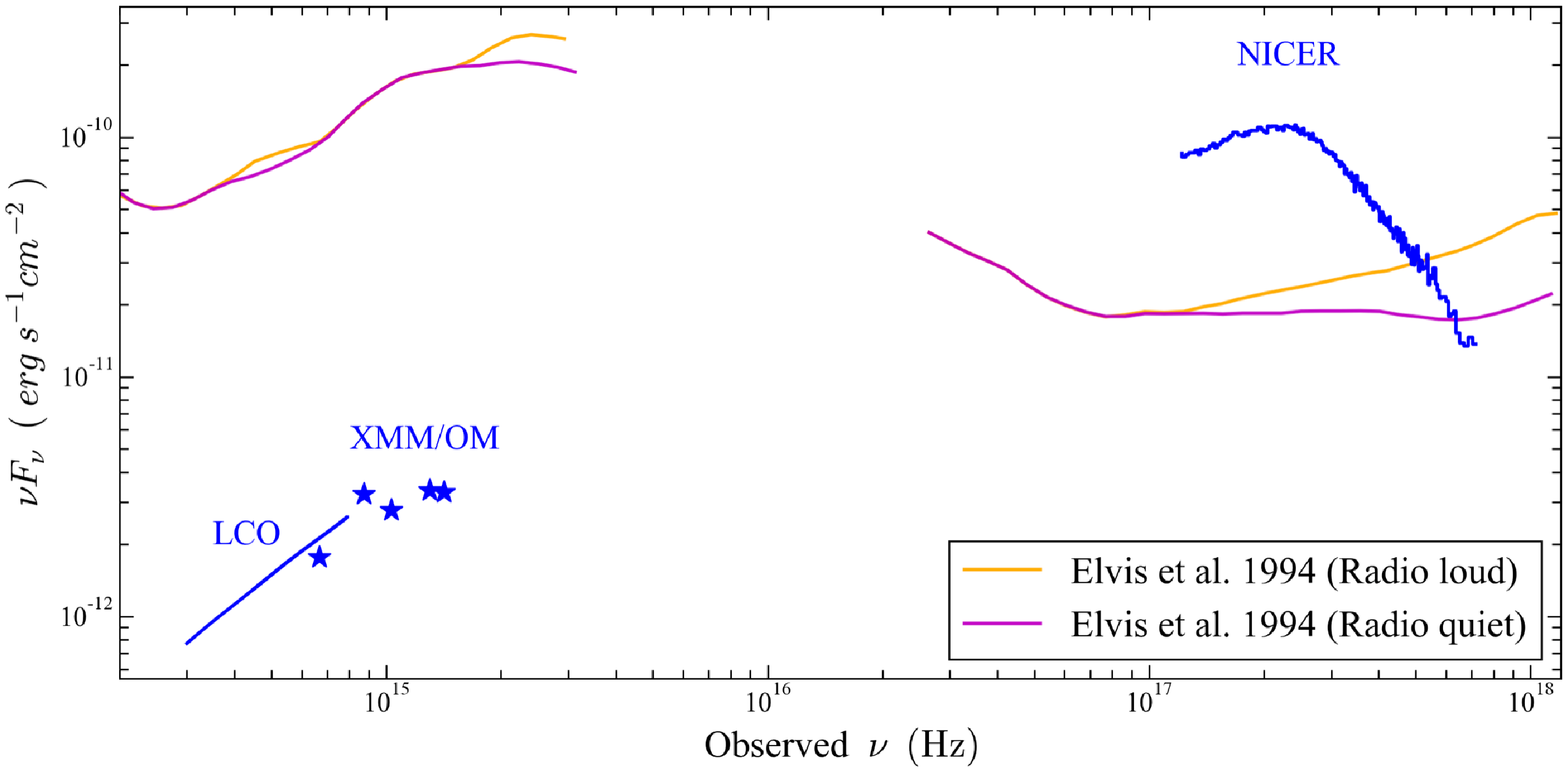}
% %
% %
 \caption{Spectral energy distribution of 1ES\,1927+654 at late times ($t\gtrsim 450$). The figure illustrates the {\it NICER} (August 3 2019, ID: 2200190277), {\it XMM-Newton}/OM (May 6 2019, ID: 0843270101) and Las Cumbres Observatory (July 25 2019) observations of 1ES\,1927+654, together with the typical spectral energy distribution of radio-quiet and radio-loud AGN\cite{Elvis:1994cr}. The host galaxy emission was subtracted from the optical/UV data (Li et al. in preparation).}
 \label{fig:latetime_SED}
\end{figure}

At times $\gtrsim 300$\,days after the event the optical/UV flux had decreased strongly (\citealp{Trakhtenbrot:2019qy}; panel\,\,D of Fig.\,\ref{fig:longtermLC}), and most of the emission could be produced in the X-rays. The variability in the X-rays and optical/UV are in fact completely uncorrelated (Fig.\,\ref{fig:longtermLC}) and a very strong emission in the X-ray band does not correspond to an increased flux in the optical/UV. While the X-ray emission oscillates between $\sim 10^{43}\rm\,erg\,s^{-1}$ and $\sim 2\times10^{44}\rm\,erg\,s^{-1}$, the bolometric luminosity inferred from the 5100\,\,$\AA$ flux (using a bolometric correction of $\kappa_{\rm bol}$=9; \citealp{Kaspi:2000zy}) is considerably lower ($\sim 5\times10^{42}\rm\,erg\,s^{-1}$). Similarly, even considering the bolometric luminosity from the near-UV ($\kappa_{\rm bol}$=2--4; \citealp{Trakhtenbrot:2012fa}) one would obtain $\sim (1-3)\times10^{43}\rm\,erg\,s^{-1}$. Thus that at the highest X-ray luminosities, the 0.5--10\,keV emission is $\sim 1-2$ orders of magnitude higher than the bolometric luminosity inferred from the optical or UV. This is also illustrated in Fig.\,\ref{fig:latetime_SED}, which shows that the spectral energy distribution of the source appears to be dominated by the X-ray emission, and is very different from what is typically found in AGN \citep{Elvis:1994cr}. Therefore at $L_{0.5-10}\simeq 10^{44}\rm\,erg\,s^{-1}$ we could consider a 0.5--10\,keV bolometric correction of $\sim 1$, and the maximum luminosity of 1ES\,1927+654 would correspond to $\lambda_{\rm Edd}\sim1$ for $M_{\rm BH}\sim 10^{6}M_{\odot}$ (see panel\,\,A of Fig.\,\ref{fig:longtermLC}). 

\end{document}